\documentclass[aps,twocolumn,showpacs]{revtex4}
\usepackage{amsmath}
\usepackage{graphicx}
\usepackage{dcolumn}
\usepackage{verbatim}
\usepackage{times}
\usepackage{subfigure}
\usepackage{bm}
\usepackage{color}
\usepackage[colorlinks,dvipdfm]{hyperref}
\usepackage{subfigure}

\begin{document}

\title{Continuous quantum phase transition between two topologically
distinct valence bond solid states associated with the same spin value}
\author{Dong Zheng$^1$, Guang-Ming Zhang$^1$, Tao Xiang$^{2,3}$, and
Dung-Hai Lee$^{4,5}$}
\affiliation{$^1$Department of Physics, Tsinghua University, Beijing, 100084, China \\
$^{2}$Institute of Physics, Chinese Academy of Sciences, Beijing 100190,
China;\\
$^{3}$Institute of Theoretical Physics, Chinese Academy of Sciences, Beijing
100190, China\\
$^{4}$Department of Physics, University of California at Berkeley, Berkeley,
CA 94720, USA \\
$^{5}$Material Science Division, Lawrence Berkeley National Laboratory,
Berkeley, CA 94720, USA}
\date{\today}

\begin{abstract}
{We propose a one-dimensional quantum Heienberg spin-2 chain, which exhibits
two topologically distinct valence bond solid states in two different
solvable limits. We then construct the phase diagram and study the quantum
phase transition between these two states using the infinite time evolving
block decimation algorithms. From the scaling relation between the
entanglement entropy and correlation length, we determine that the central
charge for the underlying critical conformal field theory is $c=2$.}
\end{abstract}

\maketitle

\section{Introduction}

Recently there have been considerable interests in the investigations of
topological ordered states, and the quantum phase transitions between them.
Since topological ordered phases usually do not exhibit conventional
symmetry breaking, these phase transitions naturally can not be described by
the conventional Landau-Ginzburg paradigm. Despite of the lack of local
order parameters, tremendous progresses have been made in characterizing
topological ordered states. Properties such as ground state degeneracy,
quasiparticle statistics, existence of edge states, topological entanglement
entropy,\cite{Kitaev2,Xiaogang} and entanglement spectrum\cite{HuiLi} have
been proposed and used to distinguish different topological ordered states.
In contrast, the study of topological phase transitions is still in its
infancy, and progresses are in demand.

One dimensional quantum spin chains have been a subject of interests for
many years. It started with the famous Haldane conjecture,\cite{Haldane}
followed by the Affleck-Kennedy-Lieb-Tasaki (AKLT) construction of the
valence bond solid (VBS) states and their associated parent Hamiltonians.%
\cite{AKLT} Moreover, the SU(2) AKLT model has been generated by introducing
q-deformed SU(2) group,\cite{zittartz} supersymmetry,\cite{arovas} and
higher symmetric groups, such as SU(n),\cite{affleck-greiter} SP(n),\cite%
{rachel}, and SO(n).\cite{Tu2} More recently a systematic method for
constructing translational invariant VBS state for the general Lie group has
been proposed.\cite{Tu}

For the $S=1$ VBS state, there is an appealing physical picture where each
spin-1 is decomposed into two ``virtual''\ spin-1/2's. Across each valence
bond, two neighboring virtual spins pair into a singlet. den Nijs and
Rommelse proposed a nonlocal string order parameter (SOP) which revealed a
hidden ``diluted antiferromagnetic order''.\cite{Nijs} Kennedy and Tasaki
found an unitary transformation that turns the nonlocal SOP to a local
ferromagnetic order parameter associated with a hidden $Z_{2}\otimes Z_{2}$
symmetry.\cite{Kennedy} However, it is extremely difficult to generalize
such a description to the cases of higher quantum integer spin chains.

In this paper, we present a model for the $S=2$ chain which exhibits two
distinct VBS states in different parameter regimes. For one of the states,
each spin-$2$ is decomposed into two virtual spin-$1$'s, and for the other
it is decomposed into two spin-$3/2$'s. These virtual spins then pair up
across every nearest neighbor bonds. In the following, we shall refer to
these two states as VBS$_{1}$ and VBS$_{3/2}$ states. For an open chain, the
ground states have 9- and 16-fold degeneracies in these two cases,
respectively. Hence these two VBS states have different topological order.
Interestingly, there is a continuous quantum phase transition between these
two VBS states. By analyzing the relation between the von Neumann
entanglement entropy and the spin-spin correlation length,\cite%
{tag,pollmann-moore} we deduce the central charge associated with the
critical conformal field theory to be two. We further conjecture that the
underlying critical field theory may be described by the level-four $SU(2)$
Wess-Zumino-Witten model.

The paper is organized as follows. In Sec. II, we review the properties of
VBS$_{1}$ and VBS$_{3/2}$ states. In Sec. III, the quantum phase transition
for the above $S=2$ spin model is explored using the infinite time evolving
decimation method.\cite{Vidal} The entanglement spectrum around the phase
transitions is studied and the central charges for the underlying conformal
field theory at the critical line are determined. A summary is given in Sec.
IV.

\section{Two distinct VBS states of a spin-2 chain}

The model Hamiltonian of the spin-2 chain is proposed as
\begin{eqnarray}
H &=&\sum_{i}\left[ J_{2}P_{2}(i,i+1)+J_{3}P_{3}(i,i+1)+P_{4}(i,i+1)\right]
\notag \\
&=&\sum_{i}\left[ \frac{189J_{3}-400J_{2}+30}{420}\left( \mathbf{S}_{i}%
\mathbf{S}_{i+1}\right) \right.  \notag \\
&&\text{ \ \ \ }-\frac{40J_{2}+7J_{3}-9}{360}(\mathbf{S}_{i}\mathbf{S}%
_{i+1})^{2}  \notag \\
&&\text{ \ \ \ }+\frac{10J_{2}-5J_{3}+1}{180}(\mathbf{S}_{i}\mathbf{S}%
_{i+1})^{3}  \notag \\
&&\text{ \ \ \ }\left. +\frac{20J_{2}-7J_{3}+1}{2520}(\mathbf{S}_{i}\mathbf{S%
}_{i+1})^{4}\right] .  \label{eq:model}
\end{eqnarray}%
where $P_{T}(i,i+1)$ is the SU(2) symmetric operator that projects the spin
states associated with sites $i$ and $i+1$ into the total spin-$T$
multiplet. The coupling constants $J_{2}$ and $J_{3}$ are all positive. In
order to make the paper self-contained, we first review the VBS$_{1}$ and VBS%
$_{3/2}$ states, respectively.

\subsection{The VBS$_{1}$ state - AKLT state}

To construct this state, we view each spin-2 as a symmetric product of two
virtual spin-$1$'s. In the VBS$_{1}$ state, two neighboring virtual spin-1
form a singlet. The direct product of the spin-2 multiplets on neighboring
sites can be decomposed into a direct sum of the total spin $S=0,1,2,3,4$
multiplets. Due to the singlet pairing of the neighboring virtual spin-1's,
the total spin $S=3,4$ multiplets can not be generated. For $J_{2}=0$, the
model Hamiltonian only penalizes the $S=3,4$ two-spin states. Hence VBS$_{1}$
state is the unique ground state of Eq.(\ref{eq:model}). In an open chain,
the unpaired virtual spin-$1$'s at the two ends are free, and they give rise
to the $3\times 3=9$ fold ground state degeneracy.\cite{AKLT,Kennedy}

In order to write down the ground state wave function, we can use the
Schwinger boson representation, and the spin-$2$ operators can be expressed
as
\begin{equation}
S_{i}^{+}=a_{i}^{\dag }b_{i},\ S_{i}^{-}=b_{i}^{\dag }a_{i},\
S_{i}^{z}=(a_{i}^{\dag }a_{i}-b_{i}^{\dag }b_{i})/2,
\end{equation}%
with a local constraint $a_{i}^{\dag }a_{i}+b_{i}^{\dag }b=4$. Then the $S=2$
AKLT VBS ground states can be expressed in a simple form,\cite{Arovas}
\begin{equation}
|\Psi _{\mathrm{AKLT}}\rangle =\prod_{i}(a_{i}^{\dag }b_{i+1}^{\dag
}-b_{i}^{\dag }a_{i+1}^{\dag })^{2}|vac\rangle   \label{EqAKLT}
\end{equation}%
where each $\left( a_{i}^{\dag }b_{i+1}^{\dag }-b_{i}^{\dag }a_{i+1}^{\dag
}\right) $ creates a singlet bond composed of two spin-$1/2$ between $i$ and
$i+1$ sites. Furthermore, by re-arranging the creation operators in Eq.(\ref%
{EqAKLT}) and combining operators with the same site operators together, $%
|\Psi _{\mathrm{AKLT}}\rangle $ can be written in a matrix product state
form straightforwardly,
\begin{equation}
|\Psi _{\mathrm{AKLT}}\rangle =\sum_{i_{1},i_{2},\cdots ,i_{N}=-2}^{2}%
\mathrm{Tr}(A^{[i_{1}]}A^{[i_{2}]}\cdots A^{[i_{N}]})|i_{1}i_{2}\cdots
i_{N}\rangle ,
\end{equation}%
where $\left\{ A^{[m]}\right\} $ with $m=\pm 2,\pm 1,0$ are $3\times 3$
matrixes,%
\begin{gather}
A^{[-2]}=\left(
\begin{array}{ccc}
0 & 0 & 0 \\
0 & 0 & 0 \\
2\sqrt{6} & 0 & 0%
\end{array}%
\right) ,\text{ }A^{[-1]}=\left(
\begin{array}{ccc}
0 & 0 & 0 \\
-2\sqrt{3} & 0 & 0 \\
0 & 2\sqrt{3} & 0%
\end{array}%
\right) ,  \notag \\
A^{[0]}=\left(
\begin{array}{ccc}
2 & 0 & 0 \\
0 & -4 & 0 \\
0 & 0 & 2%
\end{array}%
\right) ,\text{ }A^{[1]}=\left(
\begin{array}{ccc}
0 & 2\sqrt{3} & 0 \\
0 & 0 & -2\sqrt{3} \\
0 & 0 & 0%
\end{array}%
\right) ,  \notag \\
A^{[2]}=\left(
\begin{array}{ccc}
0 & 0 & 2\sqrt{6} \\
0 & 0 & 0 \\
0 & 0 & 0%
\end{array}%
\right) .
\end{gather}

\subsection{The VBS$_{3/2}$ State - SO(5) symmetric state}

Instead of splitting a spin-$2$ into two virtual spin-$1$'s, one can also
split it into two spin-$3/2$'s. In the following, we shall first view the VBS%
$_{3/2}$ state as the AKLT state of a larger symmetry group, which is
equivalent to the $SO(5)$ symmetric matrix product state in a two-leg
electronic ladder\cite{sczhang}. Afterwards we will rephrase everything in
terms of the physical spin $SU(2)$. As pointed out in Ref.\cite{Congjun},
one can view the $\pm 3/2,\pm 1/2$ states of a spin-3/2 as the four states
of the spinor representation of $SO(5)$. Similarly one can regard the $\pm
2,\pm 1,0$ states of spin-$2$ as the five-dimensional vector irreducible
representation (IR) of $SO(5)$. Analogous to decomposing a spin-$1$ vector
IR of $SU(2)$ into two virtual spin-$1/2$'s spinor IR of $SU(2)$, we can
view the vector IR as the symmetric component of the tensor product of two
virtual spinor IR's, i.e.,
\begin{equation}
\underline{4}\otimes \underline{4}=\underline{1}\oplus \underline{5}\oplus
\underline{10}.  \label{decom1}
\end{equation}%
The numerals are the dimensions of the $SO(5)$ IR's. The tensor product of
two $\underline{5}$'s on adjacent sites decomposes into
\begin{equation}
\underline{5}\otimes \underline{5}=\underline{1}\oplus \underline{10}\oplus
\underline{14}.  \label{decom2}
\end{equation}%
Comparing expressions of Eq.(\ref{decom1}) and Eq.(\ref{decom2}), we view
Eq.~(\ref{decom1}) as the tensor product of two neighboring virtual spins
after their respective partners have form $SO(5)$ singlet with other virtual
spins. Then one can find that $SO(5)$ singlet $\underline{1}$ and the
antisymmetric $\underline{10}$ appear in the decomposition but the symmetric
$\underline{14}$ is absent. Therefore, if $H=\sum_{i}P_{14}(i,i+1)$, the VBS$%
_{3/2}$ state where neighboring virtual $\underline{4}$'s pair into $SO(5)$
singlet will be the ground state. In an open chain, due to the unpaired free
$SO(5)$ spinors at two ends, the ground states are $4\times 4=16$ fold
degenerate. A clear and detailed argument of this degeneracy is given is
section III(B). Furthermore, the projection operator $P_{\underline{14}%
}(i,i+1)$ can be expressed in terms of the $SO(5)$ generators\cite{Tu2}
\begin{equation}
P_{\underline{14}}(i,j)=\frac{1}{2}\sum_{1\leq a<b\leq
5}L_{i}^{ab}L_{j}^{ab}+\frac{1}{10}(\sum_{1\leq a<b\leq
b}L_{i}^{ab}L_{j}^{ab})^{2}+\frac{1}{5}.
\end{equation}%
Because the physical spin is $SU(2)$, which is a subgroup of $SO(5)$, each
IR of $SO(5)$ must decompose into an integral number of $SU(2)$ multiplets.
Thus the $\underline{14}$ discussed above must be expressible as the direct
sum of $SU(2)$ IR obtained by decomposing the direct product of two $S=2$
multiplets. Since the 14-dimensional IR is symmetric upon the exchange of
site indices, it must only contain even-spin $SU(2)$ multiplets. A simple
calculation shows that $\underline{14}\rightarrow S=2\oplus S=4.$
Consequently, $\sum_{i}P_{14}(i,i+1)$ reduces to Eq.(\ref{eq:model}) with $%
J_{3}=0$, which is first given by Tu, Zhang, and Xiang.\cite{Tu2}

Moreover, the $SO(5)$ generators can be represented by $L^{ab}=\sum_{\alpha
,\beta }\psi _{\alpha }^{\dag }\Gamma _{\alpha \beta }^{ab}\psi _{\beta }$,
where $\psi _{j,\alpha }^{\dag }$ creates a spin-$3/2$ fermion with spin
index $\alpha =\pm 3/2,\pm 1/2$, $\Gamma ^{a},(a=1,\cdots ,5)$ are the
4-dimensional Dirac $\Gamma $ matrices, and $\Gamma ^{ab}=\frac{i}{2}[\Gamma
^{a},\Gamma ^{b}]$. By using the above representations, the VBS$_{3/2}$
state can be written as\cite{Tu2}
\begin{equation}
|\mathrm{VBS}_{3/2}\rangle =\prod_{j}\mathcal{P}_{S=2}(j)(\sum_{\alpha \beta
}\psi _{j,\alpha }^{\dag }\mathcal{R}_{\alpha \beta }\psi _{j+1,\beta
}^{\dag })|\text{vac}\rangle
\end{equation}%
where $\mathcal{P}_{S=2}(j)$ is the spin-quintet projector and $\sum_{\alpha
\beta }\psi _{j,\alpha }^{\dag }\mathcal{R}_{\alpha \beta }\psi _{j+1,\beta
}^{\dag }$ is an $SO(5)$ invariant valence bond singlet creation operator. $%
\mathcal{R}$ is the $SO(5)$ invariant matrix. This VBS$_{3/2}$ state can
also be expressed as a matrix product states (MPS)
\begin{equation}
|\mathrm{VBS}_{3/2}\rangle =\sum_{i_{1},..,i_{N}=-2}^{2}\mathrm{Tr}%
(B^{[i_{1}]}B^{[i_{2}]}\cdots B^{[i_{N}]})|i_{1}i_{2}\cdots i_{N}\rangle ,
\end{equation}%
where $\{B^{[m]}\}$ with $m=0,\pm 1,\pm 2$ are given by the following $%
4\times 4$ matrices%
\begin{gather}
B^{[-2]}=\left(
\begin{array}{cccc}
0 & 0 & 0 & 0 \\
0 & 0 & 0 & 0 \\
\sqrt{2} & 0 & 0 & 0 \\
0 & \sqrt{2} & 0 & 0%
\end{array}%
\right) ,\text{ }B^{[-1]}=\left(
\begin{array}{cccc}
0 & 0 & 0 & 0 \\
-\sqrt{2} & 0 & 0 & 0 \\
0 & 0 & 0 & 0 \\
0 & 0 & \sqrt{2} & 0%
\end{array}%
\right) ,  \notag \\
B^{[0]}=\left(
\begin{array}{cccc}
1 & 0 & 0 & 0 \\
0 & -1 & 0 & 0 \\
0 & 0 & -1 & 0 \\
0 & 0 & 0 & 1%
\end{array}%
\right) ,\text{ }B^{[1]}=\left(
\begin{array}{cccc}
0 & \sqrt{2} & 0 & 0 \\
0 & 0 & 0 & 0 \\
0 & 0 & 0 & -\sqrt{2} \\
0 & 0 & 0 & 0%
\end{array}%
\right)  \notag \\
B^{[2]}=\left(
\begin{array}{cccc}
0 & 0 & \sqrt{2} & 0 \\
0 & 0 & 0 & \sqrt{2} \\
0 & 0 & 0 & 0 \\
0 & 0 & 0 & 0%
\end{array}%
\right) .
\end{gather}

\section{Continuous quantum phase transitions}

In the model Hamiltonian Eq.(\ref{eq:model}), for $J_{2}=0$, the ground
state of the model is the VBS$_{1}$ state,\cite{AKLT} while $J_{3}=0$ it
corresponds to the VBS$_{3/2}$ state.\cite{Tu2} When both $J_{2}$ and $J_{3}$
are non-zero, the model is no longer exactly solvable. Then, we expect that
a quantum phase transition may be reached by adjusting the value of $%
J_{3}/J_{2}$. To our knowledge, this is one of the few microscopic models,
exhibiting continuous quantum phase transitions between two distinct
topological ordered phases.

\subsection{Ground state phase diagram}

In any one-dimensional quantum spin systems, the corresponding ground states
can always be simulated by the wave functions in the MPS form, as the area
law can be easily satisfied. To study the ground state properties of the
general Hamiltonian Eq.(\ref{eq:model}), we thus propose a MPS with a finite
local matrix dimension to approximate the ground state $|\psi _{g}\rangle $,
\begin{equation}
|\Psi _{g}\rangle =\sum_{\cdots m_{i}m_{i+1}\cdots }\mathrm{Tr}(\cdots
\Gamma ^{m_{i}}\Lambda \Gamma ^{m_{i+1}}\Lambda \cdots )|\cdots
m_{i}m_{i+1}\cdots \rangle ,  \label{gr}
\end{equation}%
where $m_{i}=-2,-1,0,1,2$ is the spin quantum number of $S_{i}^{z}$, $%
\left\{ \Gamma ^{m_{i}}\right\} $ is a set of $D\times D$ dimensional
matrices with bond dimension $D$ corresponding to number of states kept in
density matrix renormalization group, and $\Lambda $ is also a $D\times D$
dimensional non-negative diagonal matrix with its matrix elements $\lambda
_{\alpha }$, satisfying the normalization condition $\sum_{\alpha }\lambda
_{\alpha }^{2}=1$. The trace gives the superposition coefficients of the
Hilbert space basis. The local matrices $\left\{ \Gamma ^{m}\right\} $ and $%
\Lambda $ are set to be identical on different sites due to the translation
invariance.\cite{TransInvar}

The spirit of the infinite time evolving block decimation algorithm (iTEBDA) %
\cite{Vidal} is to do the following evolution in imaginary time
\begin{equation}
\lim_{\tau \rightarrow \infty }\frac{\exp (-H\tau )|\Psi _{0}\rangle }{%
\left\| \exp (-H\tau )|\Psi _{0}\rangle \right\| }=\lim_{N\rightarrow \infty
}\frac{\left( \exp (-H\varepsilon )\right) ^{N}|\Psi _{0}\rangle }{\left\|
\left( \exp (-H\varepsilon )\right) ^{N}|\Psi _{0}\rangle \right\| },
\end{equation}%
where $|\Psi _{0}\rangle $ is an arbitrary random initial state having
nonzero overlap with $|\psi _{g}\rangle $, $\tau $ is imaginary time, and $%
\varepsilon $ is a small interval satisfying $\varepsilon N=\tau $. It is a
projection method with high efficiency: as $\tau $ increases, the weight of
ground state in evolved state $\exp (-H\tau )|\Psi _{0}\rangle $ grows
exponentially. Next we split the Hamiltonian into two non-commutative parts,
\begin{equation}
H=\sum_{i\mathrm{=odd}}\hat{h}(i,i+1)+\sum_{i=\mathrm{even}}\hat{h}%
(i,i+1)=H_{odd}+H_{even},
\end{equation}%
where all the local two-body bond operators $\hat{h}(i,i+1)$ commute with
one another in each part $H_{odd}$ or $H_{even}$. Then Suzuki-Trotter
decomposition is used to divide the respective evolution in sequential
order,
\begin{equation}
\exp (-H\varepsilon )|\Psi _{0}\rangle =\exp (-\varepsilon H_{odd})\exp
(-\varepsilon H_{even})+O(\varepsilon ^{2})
\end{equation}%
Owing to the fact that $\exp (-\varepsilon H_{odd})$ and $\exp (-\varepsilon
H_{even})$ compose of commutating evolution operators, all the bond
evolution in each part can be implemented simultaneously, which is
compatible with the translation invariance of MPS. After each evolution
step, the domain of local matrices increases by at least one site and the
number of different local matrices increases by a factor at least $5$. Then
a singular value decomposition and bond dimension truncation are introduced
to keep the MPS in the form given by Eq.(\ref{gr}) and bond dimension fixed
in each evolution step, the detailed can be found in Ref. \cite{Vidal}.

Now we show how to calculate the ground state energy density. First we
construct the transfer matrix
\begin{equation}
G=\sum_{m=-2}^{2}\Gamma ^{m}\Lambda \otimes \left( \Gamma ^{m}\right) ^{\ast
}\Lambda  \label{transfer}
\end{equation}%
and compute its dominant eigenvalue $\eta _{1}$, as well as the associated
right and left eigenvectors $|r_{1}\rangle $ and $|l_{1}\rangle $. Then the
matrix used to compute the energy expectation value is constructed as
\begin{equation}
G_{E}=\sum_{p,q,s,t}\langle p,q|\hat{h}|s,t\rangle \Gamma ^{p}\Lambda \Gamma
^{q}\Lambda \otimes (\Gamma ^{s})^{\ast }\Lambda (\Gamma ^{t})^{\ast
}\Lambda .  \label{energytransfer}
\end{equation}%
where $\langle p,q|$ and $|s,t\rangle $ are the wave functions defined by
local Hilbert space of two adjacent sites and with the spin quantum numbers $%
p,q,s,t=-2,-1,0,1,$ $2$. The ground state energy per site $E_{g}$ can be
calculated by
\begin{equation}
E_{g}=\lim_{N\rightarrow \infty }\frac{tr\left( G^{N-2}G_{E}\right) }{%
tr\left( G^{N}\right) }=\frac{\langle l_{1}|G_{E}|r_{1}\rangle }{\eta
_{1}^{2}\langle l_{1}|r_{1}\rangle }.
\end{equation}

\begin{figure}[tbp]
\centering \includegraphics[width=3in]{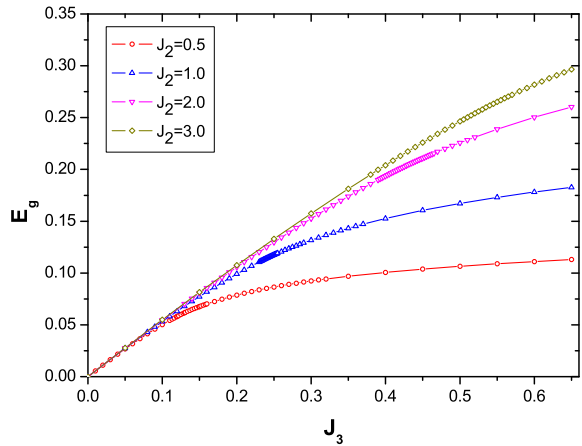}
\caption{(Color online) Ground state energy density varies as a function of $%
J_{3}$ for fixed $J_{2}$. The local matrix dimension $D$ is set to be $300$
in the calculation.}
\label{energy}
\end{figure}

\begin{figure}[tbp]
\includegraphics[width=3in]{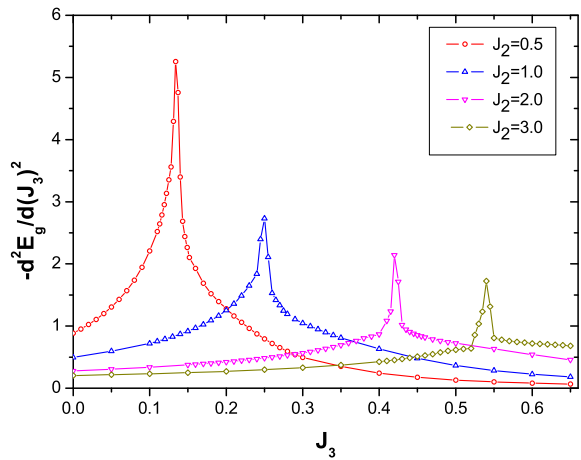}
\caption{(Color online) The second order derivative of the ground state
energy density varies as a function of $J_{3}$ for fixed $J_{2}$. }
\label{secderiv}
\end{figure}

\begin{figure}[tbp]
\includegraphics[width=3in]{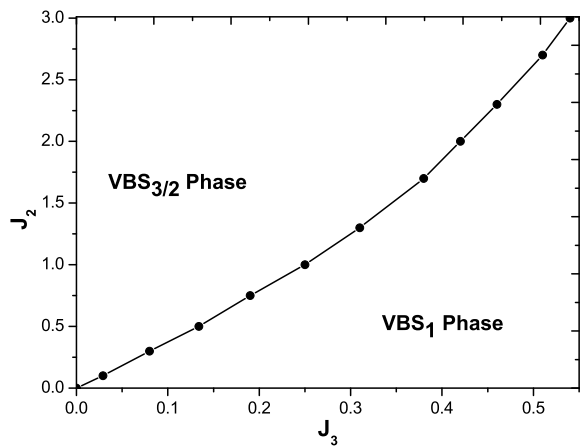}
\caption{Ground state phase diagram of the model Hamiltonian Eq.(\ref%
{eq:model}).}
\label{phasediag}
\end{figure}

For a fixed $J_{2}$, we calculate $E_{g}$ as a function of $J_{3}$. The
quantity $E_{g}$ and its first order derivative with respect to $J_{3}$ are
finite and continuous, which are displayed in FIG.\ref{energy} and FIG.\ref%
{firstderiv}. However, the second derivative of $E_{g}$ with respect to $%
J_{3}$ exhibits divergence as $J_{3}$ is tuned to a certain critical value.
This is similar to the specific heat divergence in classical phase
transitions. Such a behavior is shown in FIG.\ref{secderiv} for several
typical values of $J_{2}$. We thus conclude that the system undergoes a
second-order phase transition at zero temperature. By determining the
positions of the critical points, we thus derive the zero temperature phase
diagram in FIG.\ref{phasediag}.
\begin{figure}[tbp]
\centering \includegraphics[width=3in]{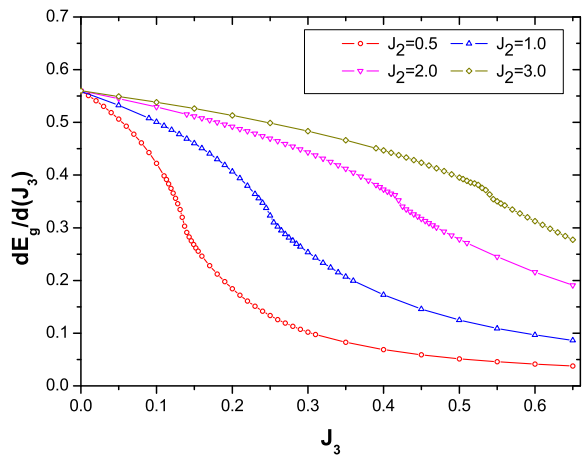}
\caption{(Color online) First derivative of the ground state energy density
varies as a function of $J_{3}$ for fixed $J_{2}$.}
\label{firstderiv}
\end{figure}

Moreover, the calculations of the spin-spin correlation length and
entanglement entropy also show a singular behavior and provide further
evidence of the second order phase transition of the model. In FIG.\ref%
{correlengthentangleentropy}, the numerical results of both spin-spin
correlation length and entanglement entropy are depicted as a function of $%
J_{3}$ for fixed $J_{2}$. At the critical point, the spin-spin correlation
length is divergent, and the entanglemnt entropy shows a cusp. Away from the
critical point, an extrapolation of correlation length and entanglement
entropy can be calculated as $D$ goes to infinity, indicating that both of
them saturate to finite values. Due to the finite spin-spin correlation
length and the form of the model Hamiltonian in terms of projection
operators, it is straightforward to prove the existence of excitation gap%
\cite{hastings} and to identify two phases separated by the critical line in
the $J_{2}-J_{3}$ phase diagram as VBS$_{1}$ and VBS$_{3/2}$ states,
respectively. The\emph{\ }computational methods of entanglement entropy and
correlation length are explained in detail in the following sections.
\begin{figure}[tbp]
\centering \includegraphics[width=3in]{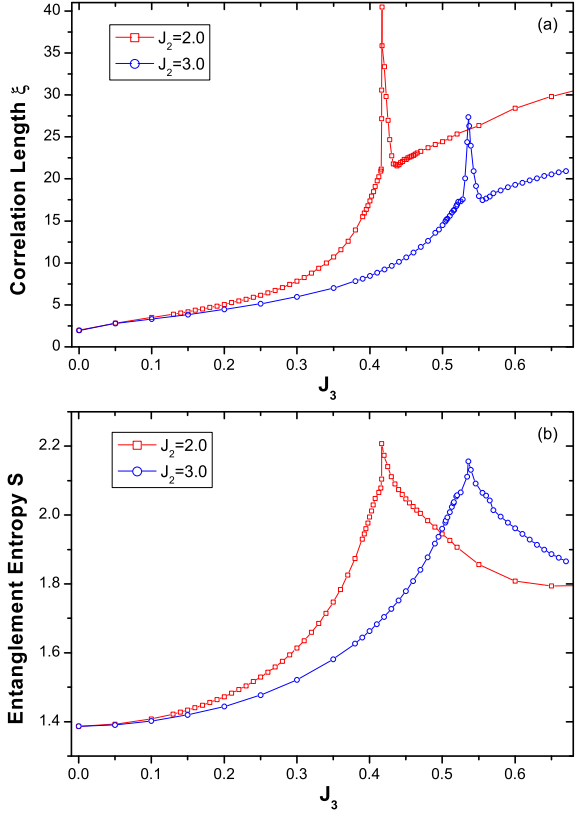}
\caption{(Color online) (a) The spin-spin correlation length varies as a
function of $J_{3}$ for two typical values of $J_{2}$. (b) The entanglement
entropy varies as a function of $J_{3}$ for two typical values of $J_{2}$.
The local matrix dimension $D$ is fixed at $300$ in this calculation.}
\label{correlengthentangleentropy}
\end{figure}

\subsection{Entanglement spectrum across the transition}

Li and Haldane\cite{HuiLi} have recently proposed that entanglement spectrum
(ES), i.e., the minus logarithms of the eigenvalues of a reduced density
matrix, can be used to characterize topological order. If there is an
entanglement gap separating the low-lying ES and the upper parts, then one
can find a one-to-one correspondence between the low-lying ES and the low
energy spectrum of individual edge excitations. In particular, the lowest
level of the entanglement spectrum for a topological ordered state should be
degenerate. When the state changes from VBS$_{3/2}$ to VBS$_{1}$ by tuning
the ratio of $J_{3}/J_{2}$, how does the topological order changes in this
process, especially when crossing the phase transition point? We will try to
use entanglement spectrum as a probe to partially answer this question.

If the MPS in Eq.(\ref{gr}) is in ``canonical form'', then upon dividing the
system into left and right parts the ground state wave function should
become
\begin{equation}
|\Psi _{g}\rangle =\sum_{\alpha }\lambda _{\alpha }|\Phi _{\alpha
}^{L}\rangle |\Phi _{\alpha }^{R}\rangle ,  \label{partition}
\end{equation}%
where $\{|\Phi _{\alpha }^{L}\rangle ,\alpha =1,2,\cdots ,D\}$ and $\{|\Phi
_{\alpha }^{R}\rangle ,\alpha =1,2,\cdots ,D\}$ are orthogonal basis states
of the left and right semi-infinite chain,
\begin{equation}
\langle \Phi _{\alpha }^{L}|\Phi _{\beta }^{L}\rangle =\langle \Phi _{\alpha
}^{R}|\Phi _{\beta }^{R}\rangle =\delta _{\alpha \beta }
\label{canonical_global}
\end{equation}%
It can be shown that the canonical condition Eq.(\ref{canonical_global})
imposes the following constraint\cite{Perez} on the $\Gamma ^{m}$ and $%
\Lambda $ in Eq.(\ref{gr}),
\begin{equation}
\sum_{m}\Gamma ^{m}\Lambda ^{2}(\Gamma ^{m})^{\dag }=\sum_{m}(\Gamma
^{m})^{\dag }\Lambda ^{2}\Gamma ^{m}=\mathbf{I}_{D^{2}\times D^{2}}.
\label{canonical_local}
\end{equation}%
In general, a MPS has a gauge freedom, because the local matrixes can be the
same up to a similarity transformation. In canonical form, it is extremely
easy and straightforward to write down the entanglement spectra $P_{\alpha
}=-\log (\lambda _{\alpha }^{2})$ . In our calculation, we perform the
canonical transformation\cite{Orus} explicitly at the end of iTEBDA and
obtain the MPS in its canonical form.
\begin{figure}[tbp]
\centering \includegraphics[width=3.2in]{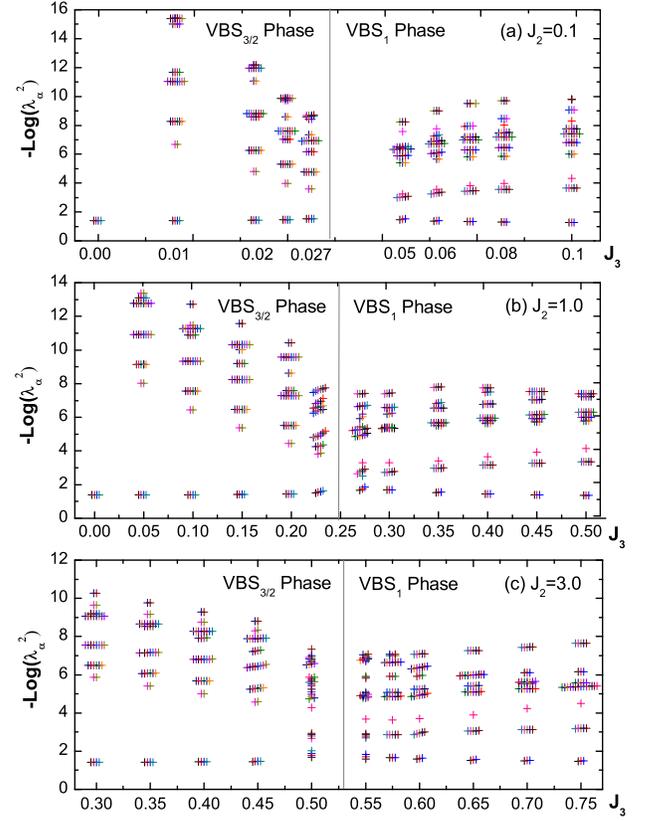}
\caption{(Color online) The thirty-sixth lowest values of the entanglement
spectra for the VBS$_{3/2}$ and VBS$_{1}$ phases on two sides of the
critical points. (a) $(J_{2},J_{3})=(0.1,0.029)$, (b) $%
(J_{2},J_{3})=(1.0,0.25)$, (c) $(J_{2},J_{3})=(3.0,0.53)$. A light gray
vertical line is put on the critical point to separate the two phases. Each
entanglement spectrum is represented by a small cross, and the degenerate
spectra are spatially staggered a little bit in horizontal direction to
distinguish and count them. In this calculations, the local matrix dimension
$D$ is set to be $300$.}
\label{EntangleSpectra}
\end{figure}

The 36 lowest values of ES are plotted in FIG.\ref{EntangleSpectra} for $%
J_{2}=0.1,1.0$ and $3.0$. It can be clearly seen that, for the VBS$_{3/2}$
state, the lowest entanglement eigenvalue is four-fold degenerate in one to
one correspondence with the four-fold degenerate edge states; while for the
VBS$_{1}$ state, the lowest eigenvalue of the ES is three-fold degenerate.
Above this degenerate eigenvalues there exists a large gap, so the
degenerate levels are protected topologically. So such a calculation of the
ES can be used to confirm that both VBS$_{1}$ and VBS$_{3/2}$ are really
topological ordered states, as well as that the ground state degeneracy of a
long enough open chain are really $16$ and $9$ respectively. When
approaching the critical points, the degeneracies are gradually lifted but
the gap still survives. However, in the vicinity of quantum critical points $%
(J_{2},J_{3})=(0.1,0.029),(1.0,0.25),(3.0,0.54)$, the degeneracies in the ES
no longer exist and the large gaps between the degenerate lowest level and
the higher levels are no longer present. Due to the expected finite-size
effect near the critical region, so far we can not simply conclude that the
topological order is destroyed completely at the critical points. Recently a
partition with a very non-local real space cut has been proposed\cite%
{Arovas2}, and some new light has shed on using ES to detect non-local
orders in gapless spin chains. However, it still needs further investigation
to clarify this question and will be done in future works.

\subsection{Central charge on the critical line}

At the critical points, the system should be described by conformal
invariant quantum field theories. For such theories the central charge
encodes information about the universality class. According to conformal
field theory, the von Neumann entanglement entropy should diverge
logarithmically with the correlation length\cite{Calabrese},
\begin{equation}
S_{e}=\frac{c}{6}\ln (\xi )+S_{0},
\end{equation}%
where $S_{e}$ is the entanglement entropy between two semi-infinite parts of
a whole chain, $c$ is the central charge, $\xi $ is spin-spin correlation
length in units of lattice spacing, and $S_{0}$ is a non-universal constant.
For a MPS in canonical form, entanglement entropy can be calculated easily%
\cite{tag,pollmann-moore}:
\begin{equation}
S_{e}=-\sum_{\alpha }\lambda _{\alpha }^{2}\ln \lambda _{\alpha }^{2}
\label{entangleentropy}
\end{equation}%
where $\lambda _{\alpha }$ are the coefficients in Eq.(\ref{partition}).

The spin correlation length can be deduced by two points spin-spin
correlation function as
\begin{widetext}
\begin{eqnarray}
\label{correlation} \lim_{\left\vert i-j\right\vert \rightarrow
\infty }\lim_{N\rightarrow \infty }\langle S_{i}^{z}S_{j}^{z}\rangle
&=&\lim_{\left\vert i-j\right\vert \rightarrow \infty
}\lim_{N\rightarrow \infty }\frac{tr\left( G^{N-\left\vert
i-j\right\vert -1}G_{z}G^{\left\vert i-j\right\vert
-1}G_{z}\right) }{tr\left( G^{N}\right) } \\
&=&\lim_{\left\vert i-j\right\vert \rightarrow \infty
}\frac{\left\vert \langle l_{1}|G_{z}|r_{1}\rangle \right\vert
^{2}}{\eta _{1}^{2}\langle l_{1}|r_{1}\rangle }+\frac{\langle
l_{1}|G_{z}|r_{2}\rangle \langle
l_{2}|G_{z}|r_{1}\rangle }{\eta _{1}\eta _{2}\langle l_{1}|r_{1}\rangle }%
\left( \frac{\eta _{2}}{\eta _{1}}\right) ^{\left\vert
i-j\right\vert } \nonumber
\end{eqnarray}
\end{widetext}where transfer matrix $G$, eigenvalues $\eta _{1}$, and
eigenvectors $\langle l_{1}|$ and $|r_{1}\rangle $ have the same definition
as in Eq.(\ref{transfer}). $\eta _{2}$ is the second largest magnitude
eigenvalue of $G$ and $\langle l_{2}|$ and $|r_{2}\rangle $ are
corresponding left and right eigenvectors. Similar to $G_{E}$ in Eq.(\ref%
{energytransfer}), $G_{z}$ is defined by
\begin{equation}
G_{z}=\sum_{p,q}\langle p|\hat{S}^{z}|q\rangle \Gamma ^{p}\Lambda \otimes
(\Gamma ^{q})^{\ast }\Lambda .
\end{equation}%
Generally, for a state without spin long-range order, $\langle
l_{1}|G_{z}|r_{1}\rangle =0$, the two points correlation function can be
written as
\begin{equation}
\lim_{\left| i-j\right| \rightarrow \infty }\lim_{N\rightarrow \infty
}\langle S_{i}^{z}S_{j}^{z}\rangle ~~\sim ~~e^{-\frac{|i-j|}{\xi }},
\end{equation}%
where the correlation length $\xi =1/\log (|\eta _{1}/\eta _{2}|)$.

\begin{figure}[tbp]
\centering \includegraphics[width=3.4in]{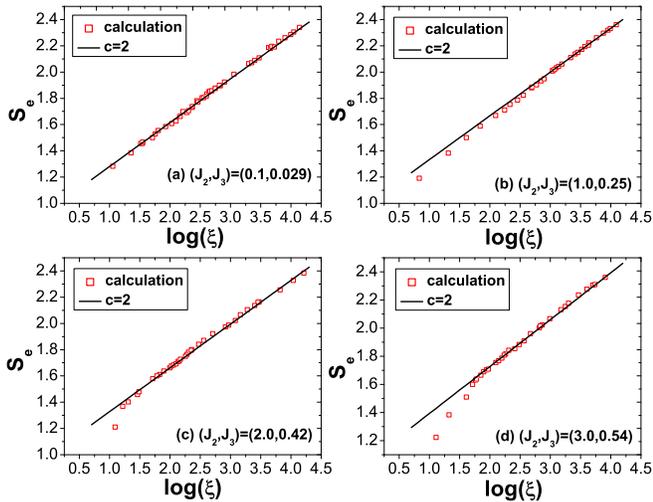}
\caption{(Color online) The scaling behavior of the entanglement entropy $%
S_{e}$ as the correlation length $\protect\xi $ in the vicinity of different
critical points. (a) $(0.1,0.029)$, (b) $(1.0,0.25)$, (c) $(2.0,0.42)$, (d) $%
(3.0,0.54)$. The black solid line is as a reference with a central charge $%
c=2$ and the red squares are numerical results. In our calculation, the
local matrix dimension increases from $12$ to $600$.}
\label{CentralCharge}
\end{figure}

Using Eq.(\ref{entangleentropy}) we have calculated $S_{e}$ and $\xi $ in
the vicinity of several different critical points on the phase boundary of
FIG.\ref{phasediag}. The associated scaling relation between $S_{e}$ and $%
\xi $ are shown in FIG.\ref{CentralCharge}. Although there may be some
deviations at small $\xi $, $S_{e}$ tends to lie on the $c=2$ line for large
$\xi $. The fact that all the central charges are approximately equal to $2$
implies a single fixed point governing the critical behavior of the entire
phase transition line. So far the conformal field theory with $c>1$ can not
be classified systematically, and therefore to determine the corresponding
conformal field theory of a fixed line with $c=2$ might be worth attempting.

Here we present some conjectures deduced from the conformal field theory
kinematics. According to the central charge value and the constituents of
the VBS$_{1}$ and VBS$_{3/2}$ states, the level-four SU(2)
Wess-Zumino-Witten model is the most possible effective field theory and the
conformal weight of the primary field given by\cite{Zamolodchikov} $\Delta
^{(j)}=j(j+1)/6$ with $j=0,1/2,1,3/2,2$. Actually, the level-four $SU(2)$
Wess-Zumino-Witten model can be regarded as the effective critical field
theory of the following spin $S=2$ antiferromagnetic Takhtajan-Babujian model%
\cite{Alcaraz}:
\begin{eqnarray}
H &=&J\sum_{i}\left[ -\frac{1}{4}+\frac{13}{48}(\mathbf{S}_{i}\mathbf{S}%
_{i+1})+\frac{43}{864}(\mathbf{S}_{i}\mathbf{S}_{i+1})^{2}\right.  \notag \\
&&\text{ \ \ \ \ }\left. -\frac{5}{432}(\mathbf{S}_{i}\mathbf{S}_{i+1})^{3}-%
\frac{1}{288}(\mathbf{S}_{i}\mathbf{S}_{i+1})^{4}\right] ,
\end{eqnarray}%
which can be written in terms of the projection operators as%
\begin{eqnarray}
H &=&J\sum_{i}\left[ J_{1}P_{1}(i,i+1)+J_{2}P_{2}(i,i+1)\right.  \notag \\
&&+J_{3}P_{3}(i,i+1)+P_{4}(i,i+1)
\end{eqnarray}%
with $J_{1}=\frac{12}{25}$, $J_{2}=\frac{18}{25}$ and $J_{3}=\frac{22}{25}$.

Compared to the model Hamiltonian Eq.(\ref{eq:model}), there appears an
additional interaction term $P_{1}(i,i+1)$ with the largest interaction
strength $J_{1}=\frac{12}{25}$. By calculating the entanglement spectrum, we
find that this critical point just lies on the boundary between the AKLT and
a dimerization phases. For the dimerization phase, the entanglement spectrum
show an even-odd difference from the topological ordered phase, i.e., if the
bipartition is done at a bond connecting left even and right odd sites, the
lowest spectrum is $5$ fold degenerate, otherwise the lowest spectrum is
non-degenerate and has a big gap with the upper part. This even-odd
difference indicates the existence of dimerization phase. The relation
between this critical point and the $SO(5)$-AKLT critical line can be
understood as follows. When we fix $J_{1}=\frac{12}{25}$ and $J_{2}=\frac{18%
}{25}$, $J_{3}$ varies from $0$ to $\frac{22}{25}$, the system evolves from
the $SO(5)$ symmetric phase to the dimerization phase, and then the AKLT
phase for $J_{3}>\frac{22}{25}$. Moreover, for the fixed value of $J_{2}$
the dimerization region shrinks and finally disappears when $J_{1}$ is
decreased, which is compatible with our ground state phase diagram.
Therefore, we expect that there exists a crossover flow from the fixed line
of the transition between the $SO(5)$-AKLT phases to the $S=2$
antiferromagnetic Takhtajan-Babujian model.

\section{Conclusion}

In summary, we propose a one-dimensional spin-2 Hamiltonian, which exhibits
two topologically distinct VBS states in different solvable limits. By using
the infinite time evolving block decimation algorithms, we have studied the
quantum phase transition between them and determined the central charge to
be $c=2$. Of course, continuous phase transition between topological phases
characterized by different number of edge states is known. For example, by
tuning the coefficient of the topological term in the $SO(3)/SO(2)$
non-linear $\sigma $ model, it is possible to induce phase transition
between VBS states associated with \textit{different spin values}. The
transition studied in this paper is very different. It takes place between
two topologically distinct VBS states associated with the \textit{same} spin
value. We are not aware of any previous study of this type of phase
transition.

\begin{acknowledgments}
The authors are grateful to Dr. Hong-Hao Tu for stimulating discussions and
earlier collaborations. We acknowledge the support of NSF of China and the
National Program for Basic Research of MOST-China. DHL was supported by DOE
grant number DE-AC02-05CH11231.
\end{acknowledgments}

\textit{Note Added} After we submitted the original version of this
manuscript for publication, a paper\cite{jiang} concerning with the similar
issue by different numerical density matrix renormalization group method on
a finite length of chain appeared on the archive, where the authors claimed
the existence of dimerization phase separated the VBS$_{1}$ and VBS$_{3}/2$
phases in the ground state phase diagram. However, if the ground state
energy density and its second-order derivative do not show any singularity
as the coupling parameters approach to the boundary of the ''dimerization
phase'' on both sides, no quantum phase transition can occur, and the claim
of dimerization phase existence is not reliable.


\begin{thebibliography}{99}
\bibitem{TransInvar} Here we just give an illustration of how to do iTEBD
numerical simulation. In actual calculations, an artificial two-fold
translational invariance is introduced in the MPS, as the local Hamiltonian
includes the nearest neighbor two-body interactions only.
\end{thebibliography}
\end{document}